\documentclass[letterpaper]{article} 
\usepackage{aaai2027}  
\nocopyright
\usepackage[hyphens]{url}  
\usepackage{graphicx} 
\graphicspath{{../}}
\usepackage[numbers,sort&compress]{natbib}
\let\originalbibliographystyle\bibliographystyle
\renewcommand{\bibliographystyle}[1]{\originalbibliographystyle{IEEEtranN}}
\AtBeginDocument{\setcitestyle{numbers,square,comma}}
\usepackage{caption}
\usepackage{algorithm}
\usepackage{algorithmic}
\usepackage{newfloat}
\usepackage{listings}
\DeclareCaptionStyle{ruled}{labelfont=normalfont,labelsep=colon,strut=off}
\floatstyle{ruled}
\newfloat{listing}{tb}{lst}{}
\floatname{listing}{Listing}
\usepackage{booktabs}
\usepackage{amsmath}
\usepackage{amssymb}
\usepackage{tikz}
\DeclareRobustCommand{\blacklabel}[1]{%
  \raisebox{-0.1em}{%
    \resizebox{!}{0.85em}{%
      \tikz{%
        \node[
          circle, fill=black, text=white,
          font=\sffamily\bfseries,
          inner sep=0.1em
        ] {#1};%
      }%
    }%
  }%
}

\title{STAGE-Med: A Multi-Agent Framework for Interactive Medical Diagnosis Based on Structured Action Spaces and Global Exploration}
\author{
\normalfont
Weiqi Li$^{1}$,
Zihao Pu$^{1}$,
Mengyun Zhang$^{2}$,
Haiyu Mao$^{1}$
}

\affiliations{
$^{1}$King's College London \qquad
$^{2}$Tsinghua University
}

\begin{document}

\maketitle

\begin{abstract}

Large language model-based multi-agent systems have demonstrated substantial potential for clinical diagnosis. However, existing approaches typically rely on predefined expert roles or group discussions. Such designs are prone to role misassignment, evidence bias, and premature diagnostic convergence. We believe that these limitations may stem from a lack of effective constraints on the diagnostic action space, which may lead agents to pursue exploratory directions that do not align with case requirements early in the diagnostic process and to propagate these initial biases into subsequent examination selection and diagnostic decision-making. To address this issue, we propose a training-free multi-agent framework that combines clinically structured action spaces with global exploration for general-purpose diagnosis. The framework comprises a diagnostic council consisting of three agents, each of which acts solely within its own action subspace whilst sharing observations and intermediate differential diagnoses with the other agents. To mitigate the risk of local optima introduced by structural constraints, an independent global agent is granted access to the complete action space and is tasked with exploring cross-modal or previously overlooked diagnostic pathways. We evaluate the proposed framework on real-world clinical cases constructed from MIMIC-IV. The results demonstrate that our approach achieves an average relative improvement of 27\% in overall diagnostic accuracy compared with state-of-the-art single-agent and multi-agent methods based on multidisciplinary consultation, group discussion, and additional exploration steps. These findings demonstrate that combining clinically structured action-space constraints with global exploration provides a more effective and efficient multi-agent coordination mechanism for general-purpose clinical diagnosis.

\end{abstract}

\section{Introduction}

Large language models (LLMs) are increasingly being explored for medical diagnosis and clinical decision support owing to their capabilities in natural language understanding, knowledge synthesis, and reasoning \cite{singhal2023clinical,hager2024limitations}. LLMs can draw on medical knowledge acquired during pretraining and adapt to downstream tasks through prompting, often without task-specific retraining. This flexibility offers the potential to alleviate the scarcity of annotated medical data and improve transferability across tasks and clinical settings \cite{moor2023foundation,singhal2023clinical}.

Building on these capabilities, LLM-based agents have been developed for diagnosis through autonomous planning, tool use, and multi-turn interaction \cite{sun2024conversational,fan2024aihospital}. In single-agent systems, an individual agent remains susceptible to uncorrected errors and reasoning biases arising from a single analytical perspective or line of reasoning \cite{li2024mediq,ferber2026autonomous}. Consequently, the researchers have proposed a multi-agent diagnostic framework. Multi-agent systems organise diagnostic reasoning through expert-role assignment and discussion among multiple agents. MedAgents and MDAgents recruit agents with medical specialist roles to analyse a case \cite{tang2024medagents,kim2024mdagents}, while Multi-Agent Conversation (MAC) uses supervisor-guided discussions among diagnostic agents \cite{chen2025enhancing}.

However, role misassignment and premature agreement on an early hypothesis can bias subsequent diagnostic actions and leave relevant evidence unexplored. Our evaluation further shows that simply increasing the number of agents or interaction steps yields limited accuracy gains at a substantially higher token cost.

We propose STAGE-Med to improve diagnostic accuracy while keeping token consumption moderate. This training-free multi-agent framework combines structured local exploration with global exploration across clinical domains, allowing agents to examine clinical details while pursuing a broader range of possible diagnoses.

STAGE-Med implements this idea through three design choices. First, we form a Diagnostic Council of three specialist agents and partition its action space according to clinical guidelines into medical history, laboratory testing, and imaging subspaces \cite{balogh2015improving}. Each specialist operates within its assigned subspace while sharing observations and intermediate differential diagnoses with the other Council members. Second, we introduce an independent Global Agent with access to the full action space to explore cross-disciplinary diagnostic pathways that individual specialists may overlook. Finally, we fuse the diagnostic results from the Council and the Global Agent to produce the final diagnosis.

We evaluate STAGE-Med on clinical cases constructed from the MIMIC-IV electronic health record (EHR) dataset \cite{johnson2023mimic}. Under the same per-agent interaction budget, it achieves 20\% higher relative diagnostic accuracy with 57\% fewer tokens than the discussion-based baseline \cite{chen2025enhancing}.

The main contributions of this work are as follows: 1) We propose STAGE-Med, a training-free multi-agent diagnostic framework that structures the Diagnostic Council's action spaces according to clinical guidelines, enables cross-action space exploration through an independent Global Agent, and fuses their diagnostic results. 2) We evaluate STAGE-Med on MIMIC-IV against single-agent and multi-agent baselines. The results show higher diagnostic accuracy than the evaluated baselines, including frameworks based on role-playing, multidisciplinary consultations, and multi-agent discussions, while keeping token consumption moderate.

\begin{figure*}[t]
    \centering
    \includegraphics[
        width=0.8\textwidth,
        height=1\textheight,
        keepaspectratio
    ]{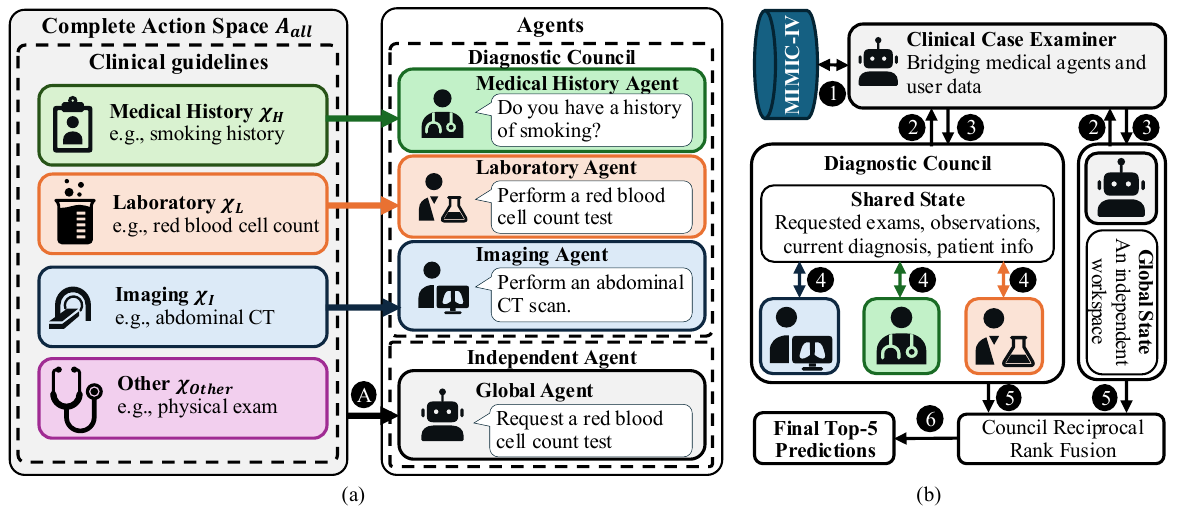}
    \caption{Overview of the action-space segmentation is shown in (a). The three agents in the Diagnostic Council are restricted to clinically structured subspaces, while the Global Agent has full access (\textbf{A}) to the complete action space. The workflow of STAGE-Med is shown in (b). Reciprocal Rank Fusion combines the Council and Global Agent's ranked diagnosis lists to obtain the five highest-ranked final predictions (Top-5).}
    \label{fig:framework}
\end{figure*}

\section{Related Work}

\paragraph{Multi-Agent Medical Reasoning.}
Multi-agent methods distribute medical reasoning among agents through expert recruitment and discussion. MedAgents recruits domain experts, collects individual analyses, and iteratively refines a shared report towards consensus; its original evaluation focuses on medical question-answering benchmarks \cite{tang2024medagents}. MDAgents selects individual or group collaboration according to task complexity \cite{kim2024mdagents}. MAC uses supervisor-guided discussions to generate diagnoses and recommend further examinations during initial and follow-up consultations \cite{chen2025enhancing}. These mechanisms organise expert participation and the aggregation of assessments.

\paragraph{Interactive Diagnostic Agents.}
Interactive diagnostic agents acquire clinical evidence through successive questions and examination requests. MediQ asks follow-up questions when it lacks confidence in a diagnostic decision \cite{li2024mediq}. MIRA combines patient dialogue with tools for requesting examinations, interpreting findings, and producing diagnostic and treatment plans \cite{ferber2026autonomous}. These systems make information acquisition part of the diagnostic process, but centre diagnostic decisions on a single agent. Coordinating evidence acquisition across diagnostic agents lies outside their scope.

\paragraph{Clinical Priors for Action Selection.}
Clinical prior knowledge can guide agents' action selection. Previous studies have incorporated clinical guidelines, safety constraints, and expert knowledge into agents for mechanical ventilation, sepsis treatment, and insulin dosing \cite{denhengst2024guideline,jia2020safe,luo2025dynamic}. In these applications, clinical knowledge helps determine which actions an agent should consider or avoid.

\section{Observations and Motivation}

\paragraph{Early Decisions Can Bias Exploration.}
Role assignment and early diagnostic hypotheses can steer subsequent actions away from relevant evidence. A role-based system recruits specialists from limited initial information; an unsuitable choice can focus later questions and tests on the wrong clinical domain. In discussion-based systems, an early shared hypothesis can similarly direct agents towards supporting evidence while alternatives receive less attention.

\paragraph{Inefficient Exploration of a Broad Action Space.}
Access to a broad range of diagnostic actions does not ensure that agents explore alternative explanations. Available actions span history questions, laboratory tests, and imaging. In a pancreatitis case from our diagnostic evaluation, the adapted MedAgents system recruited a hepatology agent after an imaging finding near the liver, while MAC pursued gastrointestinal bleeding after an abnormal blood count. Subsequent exploration followed these early interpretations. This case illustrates how an interaction budget can be spent on a narrow diagnostic direction despite a much broader set of available actions.

\paragraph{Limited Accuracy Gains from Additional Exploration.}
Increasing exploration budgets yields limited accuracy gains in our evaluated configurations. In our evaluation on MIMIC-IV, all evaluated baselines remain below 40\% overall diagnostic accuracy (Table~\ref{tab:icd_results}). For three unrestricted agents, increasing the per-agent step limit from 6 to 24 improves overall accuracy by 7.0\% relative (from 35.5\% to 38.0\%), but increases average token consumption by a factor of 4.4 (from 36,895 to 162,565 per case). Compared with the Agents $\times 3$ baseline in Tables~\ref{tab:icd_results} and~\ref{tab:token_consumption}, adding one more agent at six steps per agent improves overall accuracy by 7.9\% relative (from 35.5\% to 38.3\%), but raises token consumption by 33.3\% (from 36,895 to 49,193 per case). These results show that additional interaction and more agents alone provide modest gains at a higher token cost in this setting.

\paragraph{Motivation.}
These observations motivate guiding diagnostic actions more effectively while keeping token consumption moderate. Clinical priors can provide structure for focused exploration. Because locally constrained agents may overlook paths across subspaces, we add an independent agent with access to the full action space.

\section{Method}
As shown in Figure~\ref{fig:framework}, STAGE-Med combines a Diagnostic Council with clinically structured action spaces and an independent Global Agent with access to the full action space. The numbered markers follow the workflow in Figure~\ref{fig:framework}(b).

\subsection{Task}

We construct a clinical-case examiner agent. Unlike previous EHR-agent evaluation tasks, our agent is required to interact with the examiner agent to obtain information rather than directly retrieve information from the EHR. We model the agent's interactive differential diagnosis as a sequential information-acquisition problem. Let $D$ denote the patient's latent disease, and let $h_t$ denote the dialogue history before step $t$. At each step, the agent selects an information-acquisition action $a_t$, receives a natural-language observation $o_t$ from the examiner, and updates the history:
\begin{equation}
o_t \sim P(o_t \mid D,h_t,a_t), \qquad
h_{t+1}=h_t\cup\{(a_t,o_t)\}
\end{equation}

The objective of the diagnostic agent is to obtain clinically valuable observations under a limited information-acquisition budget and ultimately generate a differential diagnosis. Specifically, the diagnostic agent receives an admission chief complaint and demographic information, and then asks history questions or requests examinations according to the chief complaint (Figure~\ref{fig:framework}(b) \blacklabel{2}), such as asking about smoking history or requesting a lipid panel. The examiner then uses SQL to retrieve whether the EHR contains the data required by the diagnostic agent (\blacklabel{1}). If the data exist, the examiner returns them (\blacklabel{3}); otherwise, it responds, `That test was not performed.'

Unlike existing interactive diagnostic methods, the diagnostic agent we propose is designed for open-world scenarios, and the information it may request is not restricted to specific diseases or a single modality. However, the information it can obtain is limited to the real-world records in MIMIC-IV. Although this approach differs from traditional sequential information retrieval tasks, it more closely aligns with real-world clinical scenarios. Given the individual variability in patient presentations, using large language models to generate results for tests that have not been performed lacks a reliable basis and may introduce hallucinations. Therefore, we do not fill in missing tests. Furthermore, this study uses the clinical processes and final diagnostic outcomes of real-world doctors as the reference standard. Under this framework, if a test requested by the agent is not actually carried out in the real-world clinical setting, it should be regarded as an unreasonable request, and the corresponding results should not be provided.

All information output by the examiner agent and the initial chief complaint received by the diagnostic agent undergo leakage prevention (see Appendix 2). Specifically, the examiner is required to remove diagnostic names once when retrieving the data and once again when outputting the information. The results indicate that, across 500 samples, the examiner agent explicitly disclosed the diagnostic label in only 1.6\% of cases. As the aim is to simulate a real-world scenario in which patients require a diagnosis within a very short time frame, the examiner will only have access to EHR data from the patient’s first day of admission.

\subsection{Diagnostic Council and Global Agent}

\begin{figure*}[t]
   \centering
    \includegraphics[width=0.9\textwidth]{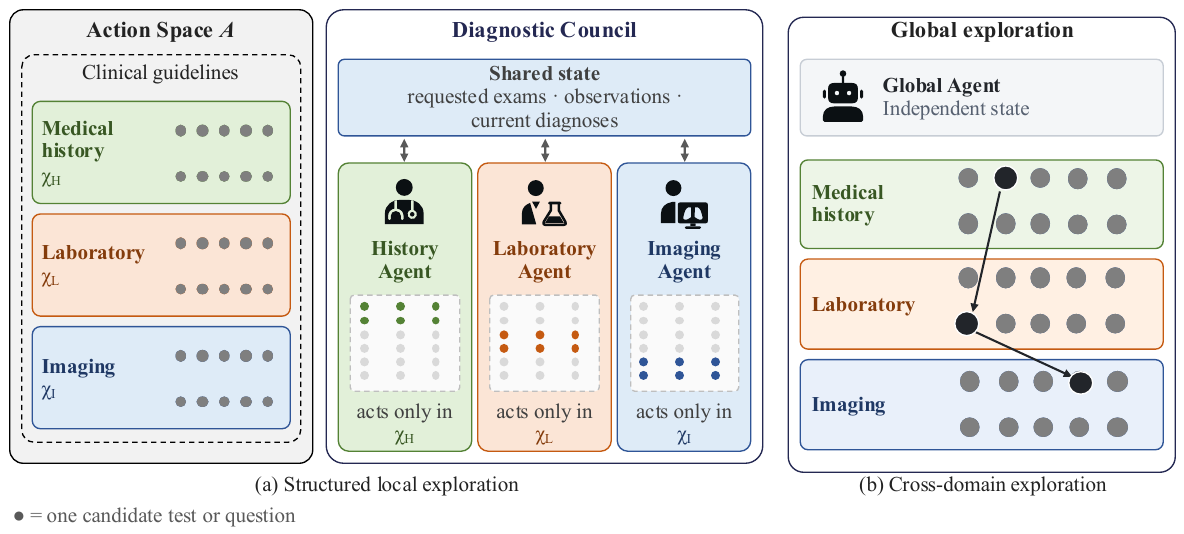}
    \caption{Complementary local and global exploration in STAGE-Med. (a) Clinical guidelines constrain each Council specialist to its assigned action subspace while the specialists maintain a shared state. (b) The Global Agent uses an independent state to explore across domains; the numbered path illustrates a possible action sequence. Dots represent questions or examinations.}
    \label{fig:framework2}
\end{figure*}

The Diagnostic Council contains three specialists whose action spaces are divided according to medical history, laboratory, and imaging evidence by prompt (Figure~\ref{fig:framework2}(a)). The criteria for dividing in this study draw upon the diagnostic process framework proposed by \citet{balogh2015improving}. This report systematically defines the basic process by which clinicians gather diagnostic information; based on this, this study has carried out a structured partitioning of the action spaces for the different agents. The spaces of information-acquisition actions available to these agents are denoted by $\mathcal{X}_H$, $\mathcal{X}_L$, and $\mathcal{X}_I$, respectively. Let $c\in\{H, L, I\}$ denote a fixed evidence-type label. Without clinical priors for action-space constraints, the agent's complete candidate action space is $\mathcal{A}^{\mathrm{all}}$. The Council's structured action space is defined as:
\begin{equation}
\mathcal{A}_C=
\bigsqcup_{c\in\{H,L,I\}}
\left(\{c\}\times\mathcal{X}_c\right)
\subseteq \mathcal{A}^{\mathrm{all}}
\end{equation}

The three agents in the Council maintain a shared state $S_t$. This state records the key information obtained and inferred during the current interaction process, including the questions and examinations already requested, $G_t$; the observations already obtained, $O_t$; the five diagnoses currently considered most likely by the system, $\hat{y}_{t,(5)}$; and the initial information $B$, namely the admission chief complaint and demographic information. Specifically,the shared state is \(S_t=(G_t,O_t,\hat{y}_{t,(5)},B)\). Therefore, the action of an agent in the Diagnostic Council is defined as:
\begin{equation}
a^{t,c}\sim\pi_{\mathrm{LLM}}\!\left(a\mid S_t,\mathcal{X}_c\right)
\end{equation}
where $\pi_{\mathrm{LLM}}$ denotes the LLM's action-selection policy. After the action is executed, $G_{t+1}$ and $O_{t+1}$ are written, and $\hat{y}_{t+1,(5)}$ is updated accordingly before entering $S_{t+1}$ (Figure~\ref{fig:framework}(b) \blacklabel{4}). At the step limit, the system returns $\hat{y}_{t,(5)}$.

The Global Agent explores independently of the Council (Figure~\ref{fig:framework2}(b)). Its state $S_t^F$ and the Council's shared state $S_t$ remain independent of each other during the consultation process. In particular, the state of the Global Agent is represented as \(S_t^F=(G_t^F, O_t^F,\hat{y}_{t,(5)}^F, B)\). The Global Agent can access the complete candidate action space $\mathcal{A}^{\mathrm{all}}$ without structured constraints. Its action is defined as:
\begin{equation}
a^t\sim\pi_{\mathrm{LLM}}\left(a\mid S_t^F\right),
\qquad a^t\in\mathcal{A}^{\mathrm{all}}
\end{equation}
After the action is executed, $G_{t+1}^F$ and $O_{t+1}^F$ are written, and $\hat{y}_{t+1,(5)}^F$ is updated accordingly before entering $S_{t+1}^F$. At the step limit, the system returns $\hat{y}_{t,(5)}^F$.

After obtaining the diagnosis lists $\hat{y}_{t,(5)}$ and $\hat{y}_{t,(5)}^F$ generated by the Council and the Global Agent, respectively (Figure~\ref{fig:framework}(b) \blacklabel{5}), we use BioLORD, a biomedical semantic representation model \cite{remy2024biolord}, to map free-text diagnosis names to their nearest-neighbor International Classification of Diseases (ICD) codes. Only the first three characters of each ICD code are retained. The system then applies Reciprocal Rank Fusion to aggregate and rank the diagnoses produced by the two pathways \cite{cormack2009rrf}. In this study, the standard parameter $k=60$ is used. For each candidate diagnosis $d$, its final fusion score is defined as
\begin{equation}
\phi_t(d)=\frac{\mathbb{I}\!\left(d \in \hat{y}_{t,(5)}\right)}{60 + r_{t,Co}(d)}+\frac{\mathbb{I}\!\left(d \in \hat{y}{^F}_{t,(5)}\right)}{60 + r_{t,G}(d)}
\end{equation}
where $r_{t, Co}(d)$ denotes the rank of candidate diagnosis $d$ in the diagnosis list generated by the Council pathway, and $r_{t, G}(d)$ denotes its rank in the diagnosis list generated by the Global pathway. Here, $\mathbb{I}[\cdot]$ is the indicator function, which is $1$ if the condition holds and $0$ otherwise.

After the fusion scores have been computed for all candidate diagnoses, the candidates are sorted in descending order according to $\phi_t$. The five candidates with the highest fusion scores are selected as the final diagnostic predictions (Figure~\ref{fig:framework}(b) \blacklabel{6}). It should be noted that, to ensure fairness, all other multi-agent baseline frameworks that do not employ a discussion-based design will also use Reciprocal Rank Fusion for the candidate diagnoses from different agents.

\section{Evaluation Setup}

\subsection{Dataset}

We construct our EHR database based on a widely used real-world EHR database, MIMIC-IV \cite{johnson2023mimic}. It is maintained by the Massachusetts Institute of Technology, and the data are from Beth Israel Deaconess Medical Center in Boston, United States. We sort the timestamps with precision to the second. To support sequential modeling, we follow the dataset construction method of \cite{liao2026agentehr}, associate events with their corresponding hospitalization records, and fill in these timestamps. We distinguish common cases from rare cases according to the occurrence frequency of diagnostic labels. Specifically, we average the weights of the diagnostic labels contained in a case to obtain the overall weight of that case, and use the mean weight of all cases as the threshold (see Appendix 7). Because this study focuses on cases with higher prevalence, we retain only cases whose weights are less than or equal to the mean. Within this sample pool, we further perform weighted random sampling according to sample weights to increase the diversity of the selected samples in the label space. Ultimately, a total of 500 samples are included in our study.

\subsection{Baseline Methods}

We select single-agent and multi-agent methods as baselines. For the single-agent baselines, we include the following methods: (1) a chief-complaint-only agent, which performs diagnosis using only the patient's chief complaint; (2) an agent based on the ReAct framework \cite{yao2023react}, which alternates reasoning and acting so that the agent can dynamically adjust subsequent decisions according to environmental feedback; (3) LLM Prune, which uses a large language model to prune the candidate action space and reduce decision complexity; and (4) MIRA \cite{ferber2026autonomous}, which introduces a patient model to simulate real physician--patient interaction and constructs a medical agent covering the complete diagnostic and treatment process. For the multi-agent baselines, we include the following three types of methods: (1) unrestricted agent teams, for which we report results with three and four agents; (2) MAC \cite{chen2025enhancing}, which improves diagnostic reasoning through supervisor-guided discussion among multiple medical agents; and (3) MedAgents \cite{tang2024medagents}, which improves diagnostic reasoning by recruiting medical agents from different disciplines to discuss the case.

\subsection{Diagnostic Agent Settings}
All agents use the same underlying large language model, Qwen2.5-72B-Instruct, and are deployed on NVIDIA A100 GPUs. For state-of-the-art baseline agents, we have deployed publicly available open-source code on local devices and adapted them for diagnostic tasks by enabling them to interact with a clinical case examiner.

Because patients have limited tolerance for lengthy clinical interactions and diagnostic examinations incur real-world costs, a diagnostic agent should reach an effective diagnosis using as few interaction steps as possible. Unless otherwise specified, the maximum number of interaction steps for each agent is therefore set to 6.
For some selected baseline agents, we additionally provide a larger interaction budget of 24 steps. This setting allows us to examine whether the baselines remain inferior to the proposed method even when they are given substantially more opportunities to collect evidence and make decisions. At each information-acquisition step, an agent asks one question or requests one examination within its permitted action space.

\subsection{Evaluation Metrics}
Following \citet{grundmann2026clinibench}, we require each agent to output its five most likely diagnoses (Top-5) in a fixed natural-language format. We then use BioLORD to map each natural-language diagnosis to its nearest ICD code \cite{remy2024biolord}. Each mapped code is subsequently truncated to its first three characters. This choice is motivated by the fact that the first three characters of both ICD-9 and ICD-10 codes define a disease category, such as acute myocardial infarction.
A prediction is considered correct if the Top-5 diagnoses contain the three-character ICD category of the ground-truth principal diagnosis, defined as the primary discharge diagnosis annotated by physicians in the original MIMIC-IV dataset \cite{johnson2023mimic}. Previous work has also used the first two characters of ICD codes for coarse-grained diagnostic comparison, as they represent broader disease groups \cite{lu2000accuracy}. Therefore, to evaluate whether an action space constrained by clinical guidelines could lead to the correct diagnostic direction, we additionally measure whether the Top-5 predictions contain the first two characters of the ground-truth ICD code. The resulting three-character and two-character ICD label spaces contain 1,910 and 238 categories, respectively. Overall Top-5 accuracy is the mean of the two match rates.

\section{Result}

\paragraph{Main Result}
At the same limit of six steps per agent, STAGE-Med achieved 13.8\% higher overall Top-5 accuracy than four unrestricted agents (43.6\% versus 38.3\%, a gain of 5.3 percentage points) and used 46.0\% fewer tokens per case (Tables~\ref{tab:icd_results} and~\ref{tab:token_consumption}). According to Table\ref{tab:icd_results}, the results show that our multi-agent architecture achieves the best performance across ICD coding tasks at different levels of granularity. Compared with the current state-of-the-art single-agent method MIRA, the multi-specialty agent framework MedAgents, and the group-discussion framework MAC, our method yields relative overall improvements of 29.0\%, 44.4\%, and 20.4\%, respectively. All results were evaluated using McNemar’s test (see Appendix 1), and all comparisons reached statistical significance at both ICD granularities (p < 0.01). These results indicate that imposing structured constraints on the action space of agents enables their actions to be more closely aligned with the current differential diagnosis, thereby reducing ineffective exploration that deviates from the case’s requirements and curbing the propagation of early diagnostic biases during subsequent information acquisition and evidence integration, ultimately enhancing the overall performance of collaborative multi-agent diagnosis.

To further analyze the effects of different constraint mechanisms, we compare our method with an LLM-based action-pruning approach. The proposed method achieves a relative overall improvement of 26.4\%. This result suggests that, in medical diagnosis, constructing the action space using clinical prior knowledge preserves clinically meaningful exploration directions more effectively than pruning actions solely according to semantic similarity. This advantage is particularly pronounced in the task with an ICD granularity of two characters. Because this task first requires the agent to identify the higher-level disease category, the correctness of the early exploration direction directly affects the subsequent diagnostic process. Clinical-prior-based action-space constraints eliminate many implausible directions at the beginning of reasoning and encourage the agent to prioritize more promising diagnostic pathways, thereby reducing errors caused by an incorrect initial direction.

STAGE-Med also achieves 14.7\% higher overall Top-5 accuracy than the three-agent baseline with 24 steps per agent (43.6\% versus 38.0\%, a gain of 5.6 percentage points), despite using six steps per agent.

\begin{table}[t]
\centering

\setlength{\tabcolsep}{3.5pt}
\renewcommand{\arraystretch}{1.02}
\begin{tabular}{@{}lccc@{}}
\toprule
\textbf{Method}
& \textbf{ICD-2}
& \textbf{ICD-3}
& \textbf{Overall} \\
\midrule
\multicolumn{4}{@{}l}{\textit{Single-Agent Methods}} \\
Chief Complaint Only
& 42.0\% & 20.8\% & 31.4\% \\
ReAct
& 47.8\% & 26.2\% & 37.0\% \\
LLM Prune
& 42.0\% & 27.0\% & 34.5\% \\
MIRA \cite{ferber2026autonomous}
& 43.2\% & 24.4\% & 33.8\% \\
\midrule
\multicolumn{4}{@{}l}{\textit{Multi-Agent Methods}} \\
MAC \cite{chen2025enhancing}
& 45.2\% & 27.2\% & 36.2\% \\
MedAgents \cite{tang2024medagents}
& 37.0\% & 23.4\% & 30.2\% \\
Agents $\times 3$
& 45.0\% & 26.0\% & 35.5\% \\
Agents $\times 3$ (steps $=24$)
& 46.4\% & 29.6\% & 38.0\% \\
Agents $\times 4$
& 46.8\% & 29.8\% & 38.3\% \\
\textbf{STAGE-Med (Ours)*}
& \textbf{53.4\%}
& \textbf{33.8\%}
& \textbf{43.6\%} \\
\bottomrule
\end{tabular}%
\caption{Performance comparison across different ICD code granularities.
ICD-2 and ICD-3 indicate whether a Top-5 prediction matches the first
two and three characters of the ground-truth principal ICD diagnosis code,
respectively.}
\label{tab:icd_results}
\end{table}

\begin{figure}[t]
    \centering
    \includegraphics[width=\columnwidth]{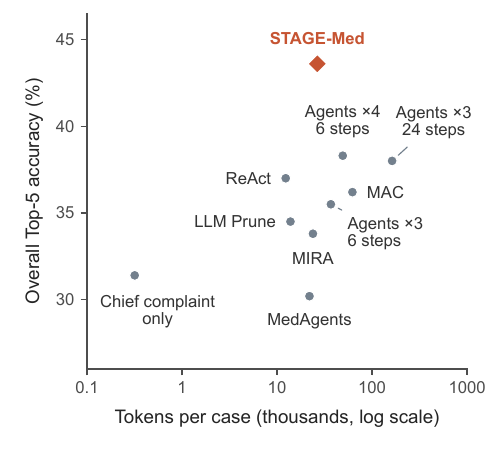}
    \caption{Overall Top-5 accuracy versus average token consumption on 500 MIMIC-IV cases. Grey circles represent the nine baseline configurations in Tables~\ref{tab:icd_results} and~\ref{tab:token_consumption}; the orange diamond denotes STAGE-Med. Steps denote the per-agent interaction limit. The token axis uses a logarithmic scale.}
    \label{fig:framework3}
\end{figure}

\paragraph{Token Consumption Analysis}

To verify that the performance gains primarily stem from the proposed architecture rather than a simple increase in token consumption, we compare the number of tokens used by each method (Figure~\ref{fig:framework3}). As shown in Table\ref{tab:token_consumption}, among the multi-agent methods, our framework consumes 21.0\% more tokens per case on average than MedAgents, the most token-efficient baseline (26,583 versus 21,962), while achieving a 44.4\% relative improvement in diagnostic accuracy. These results suggest that the improved diagnostic performance of our multi-agent framework is attributable to its collaborative design rather than merely to increased token consumption. Moreover, its relatively low token usage reduces computational overhead and facilitates practical deployment in hospital settings.

\begin{table}[t]
\centering
\small
\setlength{\tabcolsep}{2.5pt}

\begin{tabular}{@{}lrrr@{}}
\toprule
\textbf{Method}
& \textbf{Total}
& \textbf{Input}
& \textbf{Output} \\
\midrule

\multicolumn{4}{@{}l}{\textit{Single-Agent Methods}} \\
Chief Complaint Only
& 318
& 287
& 32 \\
ReAct \citep{yao2023react}
& 12,347
& 11,331
& 1,016 \\
LLM Prune
& 13,853
& 13,272
& 581 \\
MIRA \citep{ferber2026autonomous}
& 23,883
& 22,591
& 1,291 \\
\midrule
\multicolumn{4}{@{}l}{\textit{Multi-Agent Methods}} \\
MAC \citep{chen2025enhancing}
& 62,163
& 56,119
& 6,044 \\
MedAgents \citep{tang2024medagents}
& 21,962
& 18,950
& 3,012 \\
Agents $\times 3$
& 36,895
& 35,223
& 1,672 \\
Agents $\times 3$ (24 steps)
& 162,565
& 158,567
& 3,998 \\
Agents $\times 4$
& 49,193
& 46,964
& 2,229 \\
\textbf{STAGE-Med (Ours)}
& \textbf{26,583}
& \textbf{25,148}
& \textbf{1,435} \\
\bottomrule
\end{tabular}%
\caption{Average token consumption of different diagnostic methods.
Total, Input, and Output denote the average numbers
of total, input, and output tokens, respectively.}
\label{tab:token_consumption}
\end{table}

\paragraph{Ablation Study}

Given that the STAGE-Med architecture consists of multiple agents with distinct functional roles, a key question is whether each intermediate agent makes a substantive contribution to the final diagnostic performance. With up to six steps per agent, the three specialists alone achieved 38.4\% overall Top-5 accuracy, compared with 35.5\% for three unrestricted agents (Tables~\ref{tab:agent_ablation} and~\ref{tab:icd_results}). The Council’s advantage was 2.9 percentage points, or 8.2\% relative to the unrestricted team. This result suggests that imposing structured action-space constraints can provide certain benefits for diagnostic tasks. The Council’s relative advantage narrowed to 1.1\% when three unrestricted agents were given 24 steps each (38.4\% versus 38.0\%), and to 0.3\% when a fourth unrestricted agent was added at six steps per agent (38.4\% versus 38.3\%). At six steps per agent, the three unrestricted agents also outperformed the Global Agent alone (35.5\% versus 34.5\%), a gain of 1.0 percentage point or 2.9\% relative to the Global Agent. Combining the Global Agent with the three specialists achieved the highest diagnostic accuracy among all compared methods (43.6\%), a relative improvement of 13.5\% over the Council alone and 26.4\% over the Global Agent alone. This finding indicates that the performance gains of STAGE-Med do not arise solely from increasing the number of agents or from introducing structured action spaces. Instead, they depend on a balanced and complementary interaction between global unconstrained exploration and local structured collaboration. In addition, we further evaluated the diagnostic performance of each individual agent within the Council. The results show that a diagnostic agent restricted to a single sub-action space cannot achieve satisfactory diagnostic accuracy on its own. The three specialists together outperform each specialist alone when their local judgments are aggregated through collaborative reasoning.

\begin{table}[t]
\centering
\begin{tabular}{@{}lccc@{}}
\toprule
\textbf{Configuration}
& \textbf{ICD-2}
& \textbf{ICD-3}
& \textbf{Overall} \\
\midrule
Specialists Only (Council)
& 48.0\%
& 28.8\%
& 38.4\% \\
\addlinespace[2pt]
\multicolumn{4}{@{}l}{\textit{Individual Specialists}} \\
\hspace{1em}Medical History Only
& 47.8\%
& 25.8\%
& 36.8\% \\
\hspace{1em}Laboratory Only
& 35.2\%
& 20.4\%
& 27.8\% \\
\hspace{1em}Imaging Only
& 43.6\%
& 26.2\%
& 34.9\% \\
\midrule
Global Agent Only
& 43.2\%
& 25.8\%
& 34.5\% \\
\bottomrule
\end{tabular}%

\caption{Ablation results for different agent configurations. ICD-2 and
ICD-3 denote diagnostic accuracy evaluated using the first two and three
characters of the ICD codes, respectively. The indented rows represent
the individual specialists of the Council evaluated independently.}
\label{tab:agent_ablation}

\end{table}

\paragraph{Future Work}
While our current work focuses on the algorithmic design of multi-agent clinical diagnosis, deploying such systems at hospital scale introduces several important architectural challenges. Unlike conventional clinical information systems, which primarily store, retrieve, and display patient records, multi-agent systems repeatedly process and exchange large amounts of heterogeneous clinical information across multiple agents and reasoning rounds. This can increase both computation and data movement.

First, memory and storage efficiency are likely to become critical. Hospital data span electronic health records, medical imaging, and physiological monitoring, with Stanford's STARR Wave extracting approximately 75 GB of bedside-monitoring data per day \cite{callahan2023stanford}. Although not all of this information is consumed by every diagnostic case, multi-agent reasoning can repeatedly retrieve the same patient context, intermediate evidence, and model states. Consequently, the amount of data processed by the AI system can significantly exceed the amount of newly generated clinical data. Architectures that reduce redundant data movement and support efficient access to large, multimodal patient contexts may therefore provide substantial benefits \cite{he2025papi,gu2025cent,chen2025reis}.

Second, clinical deployment requires consideration of throughput and tail latency rather than single-query latency alone. A hospital-scale system may need to serve many patients concurrently, while each diagnostic case can involve several agents, multiple reasoning steps, and potentially different models. Concurrent workloads may lead to contention for compute resources, memory capacity, and memory bandwidth. Efficient scheduling, shared context management, and hardware support for concurrent agent execution therefore become important \cite{he2025papi}.

Third, energy efficiency, reliability, and data locality are particularly relevant in clinical environments. Maintaining sensitive patient data within hospital infrastructure can reduce dependence on external cloud services and improve control over privacy and availability, but it also imposes constraints on local compute capacity and power consumption.

These characteristics create opportunities for processing-in-memory (PIM) and in-storage processing (ISP). PIM can potentially reduce the movement of model parameters, embeddings, and intermediate states between memory and compute \cite{mutlu2022modernpim,he2025papi,gu2025cent}, while ISP can bring retrieval closer to large clinical data repositories \cite{chen2025reis,wang2024ndsearch}. However, clinical multi-agent workloads are heterogeneous and dynamic, combining neural inference, retrieval, structured data processing, and inter-agent communication, so only selected components may benefit from near-data processing. Determining which operations are suitable for PIM or ISP, how they should interact with CPUs and GPUs, and whether their local acceleration translates into meaningful end-to-end improvements in latency and energy efficiency therefore remains an important direction for future work. Our future study will characterize these workload behaviors and explore PIM- and ISP-based architectures for scalable, energy-efficient clinical multi-agent deployment.

\section{Conclusion}

This paper explores the key challenges facing multi-agent systems based on large language models in medical diagnosis. Existing multi-agent systems in diagnosis typically rely on predefined expert roles or open-ended group discussions, making them susceptible to errors in role allocation, early evidence bias, and premature diagnostic convergence. To address the resulting issues of inefficient exploration and the propagation of bias, we propose STAGE-Med, a multi-agent collaboration framework. This framework utilises a structured diagnostic action space organised according to clinical guidelines, enabling different agents to conduct targeted exploration within their respective constrained subspaces, whilst being supplemented by a Global Agent capable of accessing the full action space to explore cross-modal diagnostic directions. We further constructed an interactive general diagnostic task based on real electronic health records from MIMIC-IV, enabling agents to autonomously ask questions, request examinations, and update their differential diagnoses by interacting with the case examiner. Experiments demonstrate that STAGE-Med outperforms the single-agent method MIRA, the multi-agent frameworks MAC and MedAgents, and the unrestricted multi-agent baselines. Our findings highlight that combining clinically structured action space constraints with global exploration provides a more effective and efficient multi-agent coordination mechanism for general clinical diagnosis. For hospital deployment, a next step is to measure how repeated retrieval of patient records and concurrent diagnostic requests affect data movement, latency, and energy use on local hardware.

\section*{Acknowledgments}

We thank the AAAI 2027 reviewers for their constructive comments and the organisers and participants of the Arch4Health workshop at MICRO 2026 for the opportunity to discuss this work.

\bibliography{Reference}

\end{document}


\onecolumn
\section*{Supplementary Material}

\centering
\small
\setlength{\tabcolsep}{6pt}

\begin{tabular}{@{}llrrc@{}}
\toprule
Method
& Granularity
& STAGE-only
& Comparator-only
& One-sided $p$ \\
\midrule

Chief Complaint Only       & ICD-3@5 &  81 & 16 & $6.2 \times 10^{-12}$ \\
Chief Complaint  Only       & ICD-2@5 &  91 & 34 & $1.7 \times 10^{-7}$  \\
ReAct         & ICD-3@5 &  71 & 33 & $1.2 \times 10^{-4}$  \\
ReAct         & ICD-2@5 &  68 & 40 & $4.5 \times 10^{-3}$  \\
LLM Prune     & ICD-3@5 &  54 & 20 & $4.8 \times 10^{-5}$  \\
LLM Prune     & ICD-2@5 &  80 & 23 & $7.3 \times 10^{-9}$  \\
MIRA     & ICD-3@5 &  80 & 33 & $5.7 \times 10^{-6}$  \\
MIRA     & ICD-2@5 &  90 & 39 & $4.1 \times 10^{-6}$  \\
MAC  & ICD-3@5 &  68 & 35 & $7.4 \times 10^{-4}$  \\
MAC & ICD-2@5 &  74 & 33 & $4.6 \times 10^{-5}$  \\
MedAgent & ICD-3@5 &  77 & 25 & $1.2 \times 10^{-7}$  \\
MedAgent & ICD-2@5 & 101 & 19 & $5.4 \times 10^{-15}$ \\
Agent x3    & ICD-3@5 &  56 & 17 & $2.6 \times 10^{-6}$  \\
Agent x3     & ICD-2@5 &  60 & 18 & $9.9 \times 10^{-7}$  \\
Agent x3  (steps = 24)     & ICD-3@5 &  39 & 18 & $3.8 \times 10^{-3}$  \\
Agent x3 (s24)     & ICD-2@5 &  56 & 21 & $4.1 \times 10^{-5}$  \\
Agent x4         & ICD-3@5 &  36 & 16 & $3.9 \times 10^{-3}$  \\
Agent x4          & ICD-2@5 &  54 & 21 & $8.8 \times 10^{-5}$  \\

\bottomrule
\end{tabular}
\captionsetup{type=table,name=Appendix}
\captionof{table}{Pairwise comparison of STAGE-Med and the baselines using the one-sided exact McNemar test ($n=500$). STAGE-only and Comparator-only denote discordant cases correctly handled only by STAGE-Med and only by the corresponding comparator, respectively. The alternative hypothesis is $H_1$: STAGE-Med outperforms the comparator.}
\label{tab:mcnemar}

\centering
\small
\setlength{\tabcolsep}{6pt}

\begin{tabular}{@{}lrrr@{}}
\toprule
& & \multicolumn{2}{c}{Adjudicated disclosure} \\
\cmidrule(lr){3-4}
Probe set
& $n$
& Raw simulator
& After stripping \\
\midrule

Pooled across all probe modalities
& 500 & 14 (2.8\%) & 8 (1.6\%) \\

\bottomrule
\end{tabular}
\captionsetup{type=table,name=Appendix}
\captionof{table}{Diagnosis-leakage audit of the patient simulator, pooled over all $500$ probes ($100$ each for the history, laboratory, imaging, microbiology, and adversarial modalities). \emph{Adjudicated disclosure} is an LLM judge instructed to count only a literally written name of the case's ground-truth principal diagnosis, ignoring anything merely implied by findings, values, or treatments; it is reported before and after our diagnosis-term removal step. Stripping halves the disclosure rate from $2.8\%$ to $1.6\%$; excluding the adversarial probes, which are constructed to elicit a diagnosis and thus upper-bound the risk, the post-stripping rate is $4/400 = 1.0\%$. The eight remaining cases are findings whose name is itself the diagnosis (rib, scapular, and femoral fractures; aortic aneurysm; syncope), which admit no symptom-level paraphrase. }
\label{tab:leak}

\centering
\begin{minipage}{0.96\textwidth}

{\renewcommand{\lstlistingname}{Appendix}
\setcounter{lstlisting}{2}
\begin{lstlisting}[
    basicstyle=\small\ttfamily,
    caption={System prompt for the Clinical-Case Examiner Agent.},
    label={lst:appendix3},
    captionpos=b,
    numbers=none,
    frame=single,
    xleftmargin=0pt,
    framexleftmargin=6pt,
    framexrightmargin=6pt,
    framextopmargin=6pt,
    framexbottommargin=6pt,
    breaklines=true,
    breakatwhitespace=true,
    columns=fullflexible,
    keepspaces=true,
    showstringspaces=false,
    aboveskip=6pt,
    belowskip=6pt
]
You are the patient AND the hospital RESULTS SYSTEM during a
diagnostic workup. The doctor will either ORDER a test
(e.g. "Order a CBC", "Order a CT abdomen/pelvis", "Obtain a
blood smear") or ask a history/symptom question.

- ORDERED TEST: If that test WAS performed for this patient
  (its data is in the case information below), report its
  result as if the patient just went and had it done.

  For laboratory tests, give the numeric values and flags.

  For imaging, give ONLY the objective DESCRIPTIVE FINDINGS,
  including locations, sizes, and densities (e.g. "5 cm
  lobulated mass in liver segment VI"). NEVER provide the
  radiologist's IMPRESSION or conclusion. Omit phrases such as
  "consistent with", "suspicious for", and "diagnostic of",
  as well as all disease names.

  MATCH ORDERS FLEXIBLY by body region, organ, or test category,
  NOT by exact modality or wording. An order to image a region
  (e.g. "CT abdomen/pelvis" or "image the liver") is satisfied
  by ANY available imaging of that region, including CT,
  ultrasound, or MRI. Report the available objective findings.

  A laboratory order is satisfied by the relevant available
  laboratory values.

  Reply "That test was not performed." ONLY when NO related
  data exists for that body region or test category.

- HISTORY/SYMPTOM QUESTION: Answer as the patient in plain,
  everyday language. Describe only the patient's symptoms,
  experiences, events, and treatments.

ABSOLUTE RULE: Never state, name, or imply a disease or
diagnosis. Provide only objective test findings and the
patient's lay description of their experience.
\end{lstlisting}
}
\end{minipage}

{
\renewcommand{\lstlistingname}{Appendix}
\setcounter{lstlisting}{3}

\begin{lstlisting}[
    caption={Redaction filter prompt for removing diagnostic leak.},
    label={lst:appendix4},
    captionpos=b,
    basicstyle=\small\ttfamily,
    numbers=none,
    frame=single,
    xleftmargin=0.04\textwidth,
    xrightmargin=0.04\textwidth,
    framexleftmargin=6pt,
    framexrightmargin=6pt,
    framextopmargin=6pt,
    framexbottommargin=6pt,
    breaklines=true,
    breakatwhitespace=true,
    columns=fullflexible,
    keepspaces=true,
    showstringspaces=false,
    aboveskip=6pt,
    belowskip=6pt
]
You are a redaction filter for a diagnostic exam. Remove EVERY
named medical diagnosis or disease name from the text
(e.g. "hepatocellular carcinoma", "HCC", "pneumonia",
"leukemia", "myocardial infarction", "sepsis", "cirrhosis").

KEEP all objective findings, including lab values, measurements,
sizes, locations, organ names, symptoms, vital signs,
procedures, and the patient's lay descriptions.

Replace a diagnosis term with a neutral objective phrase
(e.g. "a mass" or "a procedure on the liver"), or simply delete
the diagnosis term.

Output ONLY the cleaned text.
\end{lstlisting}
}

{
\renewcommand{\lstlistingname}{Appendix}
\setcounter{lstlisting}{4}

\begin{lstlisting}[
    caption={Prompt for Global agent.},
    label={lst:appendix5},
    captionpos=b,
    basicstyle=\small\ttfamily,
    numbers=none,
    frame=single,
    xleftmargin=0.04\textwidth,
    xrightmargin=0.04\textwidth,
    framexleftmargin=6pt,
    framexrightmargin=6pt,
    framextopmargin=6pt,
    framexbottommargin=6pt,
    breaklines=true,
    breakatwhitespace=true,
    columns=fullflexible,
    keepspaces=true,
    showstringspaces=false,
    aboveskip=6pt,
    belowskip=6pt
]
WORKUP SO FAR:
{tr}

You are an INDEPENDENT diagnostician conducting your own
workup. ORDER the SINGLE most useful NEXT test, including any
laboratory test, imaging study, or culture, or ask focused
history question.

Do NOT repeat anything that has already been done.

Reply with ONLY the order or question.
\end{lstlisting}
}

{
\floatname{listing}{Appendix}
\setcounter{listing}{5}

\noindent\textbf{Diagnostic Council agent prompt}

\begin{lstlisting}[
    basicstyle=\small\ttfamily,
    numbers=none,
    frame=single,
    xleftmargin=0.04\textwidth,
    xrightmargin=0.04\textwidth,
    framexleftmargin=6pt,
    framexrightmargin=6pt,
    framextopmargin=6pt,
    framexbottommargin=6pt,
    breaklines=true,
    breakatwhitespace=true,
    columns=fullflexible,
    keepspaces=true,
    showstringspaces=false,
    aboveskip=6pt,
    belowskip=6pt
]
You are the {action space} on a diagnostic team.

CONSULTATION SO FAR:
{transcript}

CURRENT DIFFERENTIAL:
{diff}

{instr}

Reply with ONLY the order or history question
(e.g. "Order a CBC"), nothing else.
\end{lstlisting}

\vspace{1em}

\noindent\textbf{Action-space-specific instructions}

\vspace{0.4em}

\begin{center}
\small
\setlength{\tabcolsep}{8pt}
\renewcommand{\arraystretch}{1.25}

\begin{tabular}{@{}p{0.15\textwidth}p{0.75\textwidth}@{}}
\toprule
Action space & Instruction \\
\midrule

\textbf{Medical History}
&
Ask the SINGLE most useful HISTORY question concerning symptoms,
timeline, or past medical history.
\\

\textbf{Lab}
&
ORDER the SINGLE most discriminating laboratory test. Phrase it
as an ORDER, e.g. ``Order a complete blood count with
differential.'' Order the test; do NOT ask for ``information.''
\\

\textbf{Imaging}
&
ORDER the SINGLE most useful imaging study, e.g. ``Order a CT
abdomen/pelvis.''
\\

\bottomrule
\end{tabular}
\end{center}

\vspace{1em}

\noindent\textbf{Differential-generation prompt}

\begin{lstlisting}[
    basicstyle=\small\ttfamily,
    numbers=none,
    frame=single,
    xleftmargin=0.04\textwidth,
    xrightmargin=0.04\textwidth,
    framexleftmargin=6pt,
    framexrightmargin=6pt,
    framextopmargin=6pt,
    framexbottommargin=6pt,
    breaklines=true,
    breakatwhitespace=true,
    columns=fullflexible,
    keepspaces=true,
    showstringspaces=false,
    aboveskip=6pt,
    belowskip=6pt
]
Given this consultation so far, list the current top
differential diagnoses as a short, comma-separated list.

{transcript}

DIFFERENTIAL:
\end{lstlisting}

\captionof{listing}{
Prompts and action-space-specific instructions for diagnostic council
agent.
}
\label{lst:appendix6}

}

\subsection {Appendix 7: EHR Dataset Construction and Sampling}
\label{app:ehr-dataset-construction}

We construct the electronic health record dataset from
MIMIC-IV, a large-scale, real-world clinical database jointly
developed by the Massachusetts Institute of Technology and Beth
Israel Deaconess Medical Center in Boston, Massachusetts, USA. 
All clinical events are organized chronologically,
and their timestamps are standardized to a resolution of one second.
Following the data-construction procedure introduced in
, we associate each clinical event
with its corresponding hospitalization record and complete the
temporal information required for sequential modeling.

To control the distribution of diagnostic labels during sample
selection, we assign each candidate case a frequency-aware sampling
weight. For a case \(S\) associated with a set of diagnostic labels
\(Y^{*}\), the sampling weight is defined as the average inverse
frequency of its labels:

\begin{equation}
w_S
=
\frac{1}{\lvert Y^{*} \rvert}
\sum_{y \in Y^{*}}
\frac{1}{\operatorname{Count}(y)},
\label{eq:frequency-aware-weight}
\end{equation}

\noindent
where \(\operatorname{Count}(y)\) denotes the total number of
occurrences of diagnostic label \(y\) in the candidate dataset. Under
this definition, cases containing less frequent diagnostic labels
receive larger weights, whereas cases dominated by more prevalent
labels receive smaller weights.

We calculate the mean case weight, denoted by \(\bar{w}\), across the
complete candidate set and use it to divide the cases into common and
rare pools. Specifically, cases satisfying

\begin{equation}
w_S \leq \bar{w}
\label{eq:common-case-condition}
\end{equation}

\noindent
are assigned to the common-case pool, whereas cases satisfying
\(w_S > \bar{w}\) are assigned to the rare-case pool.

Because this study primarily investigates relatively prevalent
clinical cases, we retain only samples from the common-case pool. We
then perform weighted random sampling within this pool, with each
case's selection probability determined by its frequency-aware weight
\(w_S\). Consequently, cases with comparatively less frequent
diagnostic-label patterns are more likely to be selected than cases
with highly prevalent patterns.

This sampling strategy preserves the intended emphasis on common
clinical cases while improving diagnostic-label diversity within the
evaluation cohort. The resulting evaluation set contains 500 cases.

\subsection {Appendix 8: Use of Data}
\label{Use of Data}

We conducted our experiments using MIMIC-IV, a de-identified electronic health record dataset collected by Beth Israel Deaconess Medical Centre and published via PhysioNet. Authorised users may access this dataset only after completing the necessary human subjects research training and signing the PhysioNet Data Use Agreement. MIMIC-IV is a retrospective, de-identified clinical dataset. The original collection and sharing of MIMIC-IV were reviewed and approved by the Institutional Review Board (IRB) of Beth Israel Deaconess Medical Centre, and were exempt from informed consent. All researchers accessing the data have completed the necessary training, obtained certified access via PhysioNet, and agreed to comply with the PhysioNet Data Use Agreement. We have made no attempt to identify any patients, healthcare providers or institutions involved in the dataset, and no identifiable patient information is reported in this article.

\subsection {Appendix 9: Use of Generative Artificial Intelligence}
\label{Use of Ai}

Generative artificial intelligence tools were used solely to improve the grammatical accuracy, clarity, and readability of the authors’ original text. These tools were not used to generate scientific claims, experimental results, or references. All AI-assisted revisions were reviewed and verified by the authors.